\documentclass[5p,twocolumn,times,number]{elsarticle}
\pdfoutput=1

\usepackage{graphicx}
\usepackage{amsmath}
\usepackage{multirow}
\usepackage{textcomp}

\begin{document}

\begin{frontmatter}

\title{Operation of a double-phase pure argon Large Electron Multiplier Time Projection Chamber: comparison of single and double phase operation}

%\author[add1]{A.~Badertscher}
%\author[add1]{L.~Knecht}
%\author[add1]{M.~Laffranchi}
%\author[add1]{D.~Lussi}
%\author[add1]{A.~Marchionni}
%\author[add1]{G.~Natterer}
%\author[add1]{P.~Otiougova}
%\author[add1]{F.~Resnati}
%\author[add1]{A.~Rubbia}
%\author[add1]{T.~Viant}

\author{A.~Badertscher}
\author{L.~Knecht}
\author{M.~Laffranchi}
\author{D.~Lussi}
\author{A.~Marchionni}
\author{G.~Natterer}
\author{P.~Otiougova}
\author{F.~Resnati}
\author{A.~Rubbia}
\author{T.~Viant}

%\address[add1]{Institute for Particle Physics, ETH Zurich, 8093 Zurich, Switzerland}
\address{Institute for Particle Physics, ETH Zurich, 8093 Zurich, Switzerland}

\begin{abstract}
We constructed and operated a double phase (liquid-vapour) pure argon
Large Electron Multiplier Time Projection Chamber (LAr LEM-TPC)
with a sensitive area of 10x10~cm$^2$ and up to 30~cm
of drift length. The LEM is a macroscopic hole electron multiplier
built with standard PCB techniques: drifting electrons are extracted
from the liquid to the vapour phase and driven into the holes of the
LEM where the multiplication occurs. Moving charges induce a
signal on the anode and on the LEM electrodes. The orthogonally segmented upper
face of the upper LEM and anode permit the reconstruction of X-Y
spatial coordinates of ionizing events. The detector is equipped with a
Photo Multiplier Tube immersed in liquid for triggering the ionizing
events and an argon purification circuit to ensure long drift
paths. Cosmic muon tracks have been recorded and further characterization of
the detector is ongoing. We believe that this proof of principle
represents an important milestone in the realization of very large,
long drift (cost-effective) LAr detectors for next generation neutrino
physics and proton decay experiments, as well as for direct search of
Dark Matter with imaging devices.
\end{abstract}


\end{frontmatter}

\section{Introduction}
The detector described in this paper is a pure liquid argon based Time
Projection Chamber (TPC) capable of charge amplification by means of two
stages Large Electron Multiplier (LEM) with anode readout in the argon
vapour region above the liquid surface.

This detector, that we call double phase (liquid-vapour) pure argon
LEM-TPC, is a tracking and a calorimetric device that allows to
reconstruct three-dimensionally the position and the morphology of the
ionizing event in addition to the energy lost in the active volume.

The use of the LEM is motivated by the fact that it can operate in
pure argon gas and in cryogenic environment, as required in liquid
argon TPCs, moreover the gain can be adjusted to match a wide spectrum
of physics requirements.
We believe that this technology is scalable up to very large detector,
instrumenting the surface with a collection of $\sim$1x1~m$^2$
independently operating pieces: applications
of such a detector are next generation neutrino physics, proton decay
experiments~\cite{Rubbia:2004} and direct dark matter search with
imaging detectors~\cite{Rubbia:2006}.

The principle of operation of the LEM-TPC is described in
section~\ref{sec:operationPrinciple}, the construction details are
addressed in section \ref{sec:detectorDescription} and
\ref{sec:puritySystemDescription}. In section
\ref{sec:LEM-TPCOperation} we will describe the operation of the
LEM-TPC with the LEMs completely immersed in liquid argon (gain~1)
and the operation in double phase conditions (gain~10), focusing on
the difference between these two modes of operation.

\section{Principle of operation}
\label{sec:operationPrinciple}
\begin{figure}[t]
\centering
\includegraphics[width=0.95\linewidth]{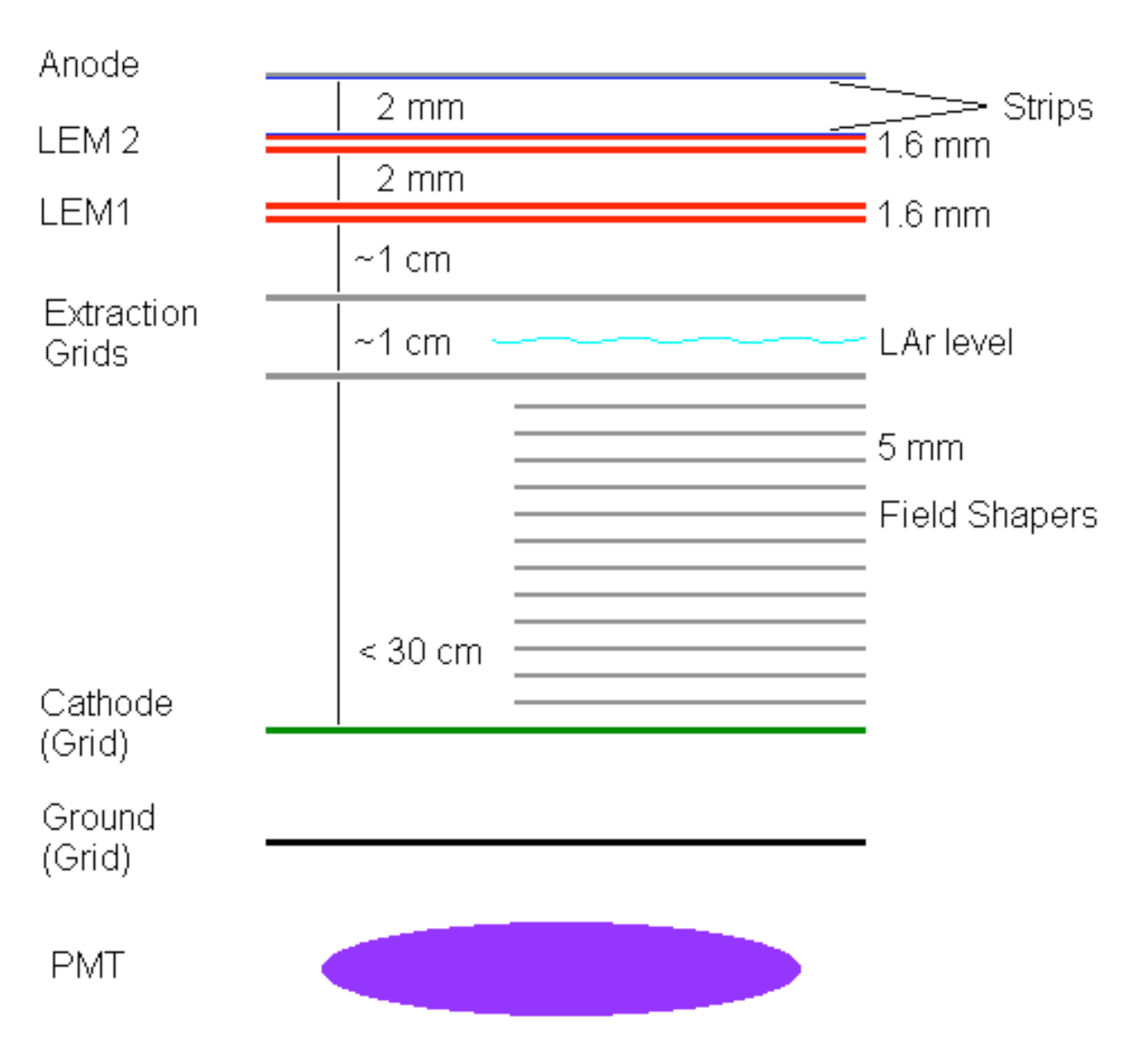}
\caption{Schematic representation of the LEM-TPC.}
\label{fig:LEM-TPCScheme}
\end{figure}

A charged particle crossing the liquid argon active volume produces
electron-ion ionization pairs; the quasi-free electrons surviving
recombination drift under the action of a
uniform electric field towards the liquid-vapour interface, see
figure~\ref{fig:LEM-TPCScheme}. Two grids across the liquid surface
provide an electric field high enough to extract electrons from the
liquid to the vapour phase~\cite{Gushchin:1981}. The electrons are
driven into the holes of the first LEM where they undergo Townsend
multiplication~\cite{Sauli:1977}; the amplified charge enters into a second
stage of multiplication and finally is collected on the anode and top
LEM face electrodes.

The LEM is a thick and macroscopic hole multiplier directly
extrapolated from the GEM detectors~\cite{Sauli:1997}. Compared to
GEMs, LEMs are more rigid and they are produced with
standard Printed Circuit Board (PCB) method: millimeter-size holes are
precisely drilled through a double side copper-cladded millimeter-thick
Flame Retardant~4 (FR4) plate. Table~\ref{tab:LEMSpecs} summarizes
the technical characteristics of the LEMs that were used.

A naturally-confined uniform electric field is obtained inside the
holes of the LEM by applying a potential difference across the two
metalized faces. This suggests to use the same formalism of the
parallel plate chamber, so that we can write the gain as $G =
\exp(\alpha x)$~\cite{Badertscher:2008}, where $x$ is the effective
amplification path length and $\alpha$ is the first Townsend
coefficient. The dependence of $\alpha$ on the gas density ($\rho$)
and the electric field ($E$) is approximated by~\cite{Aoyama:1985}:
$\alpha = A \rho \exp(-B\rho/E)$, where $A$ and $B$ are parameters
depending on the gas.

\begin{table}
\centering
  \begin{tabular}{lr}
    \hline
    \multicolumn{2}{c}{LEM specifications} \\
    \hline
    material & copper-cladded FR4\\
    active area & 10x10~cm$^2$ \\
    PCB thickness & 1.6~mm \\
    copper layer thickness & 18~$\mu$m \\
    holes diameter & 500~$\mu$m \\
    hole rim & 50~$\mu$m \\
    hole pitch & 800~$\mu$m \\
    readout strip width & 6~mm \\
    total \# readout strips & 2x16 \\
    \hline
  \end{tabular}
  \caption{Large Electron Multiplier (LEM) characteristics.}
  \label{tab:LEMSpecs}
\end{table}

With the usually adopted electric fields, about
half of the multiplied electrons coming out from the LEMs holes are
collected on the anode and half on the top face of the second LEM,
inducing signal on both planes. Additionally moving positive ions
contribute to the signal of the LEM plane.
Each electrode is divided into 16
strips 6~mm wide in order to reconstruct the position of the ionizing
event in the plane (X-Y) parallel to the liquid argon level. The depth
of the event is proportional to the drift time and the interaction
time is given by the argon scintillation recorded with a Photo
Multiplier Tube (PMT), located in the liquid below the cathode.

For a more precise description of the LEM-TPC working principle and
the description of the multiplication mechanism in the LEM holes
see~\cite{Badertscher:2008}.

\section{Detector description}
\label{sec:detectorDescription}
The aim of this section is to briefly present the detector. The
interested reader can find a complete description of the experimental
apparatus in~\cite{Badertscher:2008}. In section
\ref{sec:puritySystemDescription} we shortly discuss the liquid argon
purification system, not described in~\cite{Badertscher:2008}.

\begin{figure}[t]
\centering
\includegraphics[width=0.95\linewidth]{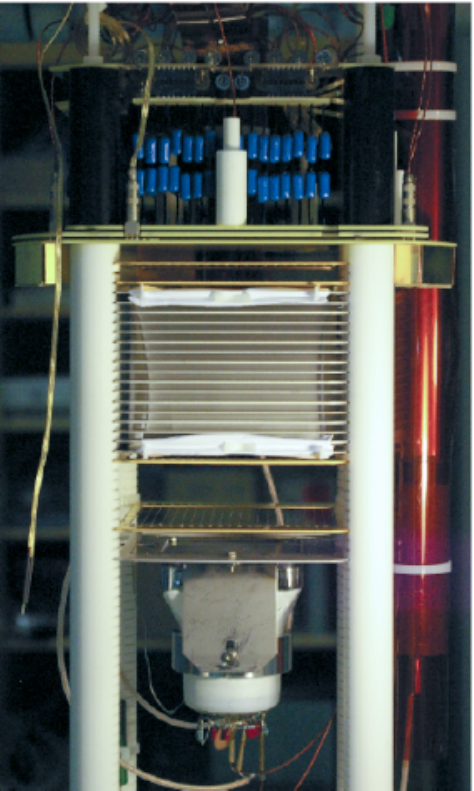}
\caption{Picture of LEM-TPC, see text.}
\label{fig:LEM-TPCPicture}
\end{figure}

A picture of the LEM-TPC is shown in
figure~\ref{fig:LEM-TPCPicture}. A cryogenic PMT~\cite{PMT}, with the
photocathode coated
with tetraphenylbutadiene (TPB) wavelength shifter, is installed below
the cathode grid, shielded by a ground grid between the PMT and the
cathode. The uniformity of the electric field for the 10~cm drift
(extensible up to 30~cm) is provided by a set of stainless steel field
shapers. The liquid argon level is between the two
extraction grids (stainless steel wires, 100~$\mu$m diameter and 5~mm
pitch) and it is monitored with three custom made capacitive level
meters. Above the
grids there are the LEM stages (LEMs characteristics are summarized in
table~\ref{tab:LEMSpecs}) and the signal collection system (called anode).
Ceramic capacitors (two 470~pF capacitor in series per channel) are
used to decouple the signal from the high voltage,
surge arresters towards ground protect the electronics from
discharges. A custom made Kapton flex-print is used to connect
the outside electronics.
The detector, constructed with UHV standards to
ensure the purity of the argon, is contained in a vessel immersed
in an external liquid argon bath to maintain stable thermodynamic
conditions. 
The digitizing and the data acquisition system is a
CAEN~SY2791~\cite{CAEN} operating at 2~MS/s. The preamplifiers
are low-noise JFET charge preamp and shaper custom developed by us
($\sim$11~mV/fC, rise-time~$\sim$0.6~$\mu$s and
fall-time~$\sim$2~$\mu$s) and directly pluggable into the CAEN
module.

\section{Purity system}
\label{sec:puritySystemDescription}
\begin{figure}[t]
\centering
\includegraphics[width=0.95\linewidth]{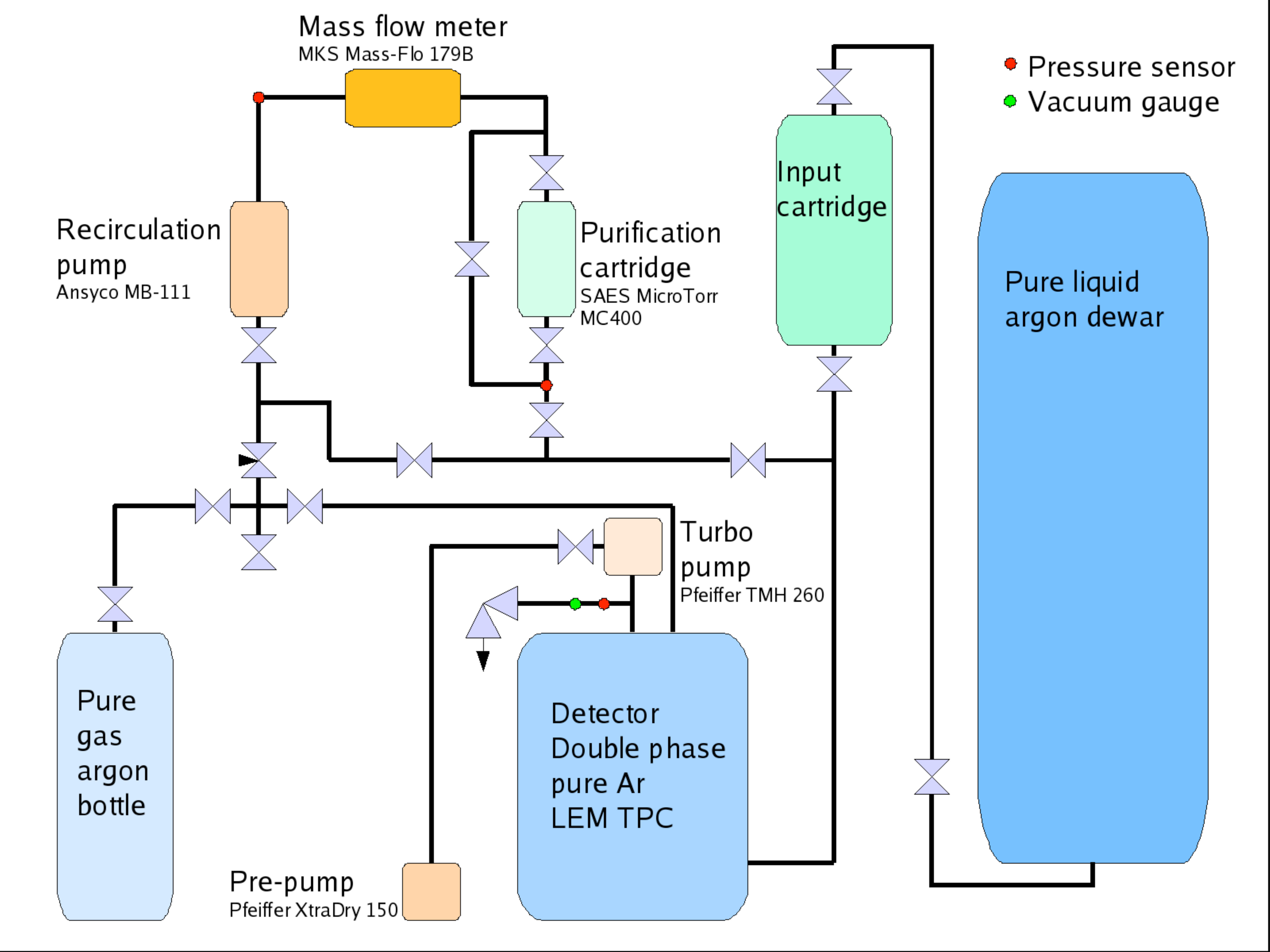}
\caption{Scheme of the purification system.}
\label{fig:recirculationScheme}
\end{figure}

Figure~\ref{fig:recirculationScheme} shows a schematic of the
liquid argon purification circuit. Two different purification systems
are used to guarantee the purity needed to drift electrons in liquid
argon: a custom made cartridge to purify the liquid
argon when filling the detector and a gas argon recirculation
system.\\
During the filling liquid argon is fed into the custom made filter,
filled with pure copper powder. The copper oxidizes trapping oxygen
impurities from the liquid; this cartridge can be regenerated by
increasing the temperature and flushing argon-hydrogen mixture to
induce CuO$_2$+2H$_2$$\rightarrow$Cu+2H$_2$O.\\
To maintain the purity for long periods the liquid argon is evaporated
and pushed by a metal bellow pump through a commercial
getter~\cite{SAES}.
The cleaned gas argon is re-condensed into the detector by means of a
spiral radiator. The argon flow is regulated via a needle valve at
the input of the pump and it is monitored with a mass flow meter
mounted between the pump and the getter.
We operate the recirculation at 7.5~slm: two days are needed for a
full volume change.

Other precautions are used to ensure a good purity:
after the assembly, the detector, the recirculation system and
the cartridges are pumped down to $10^{-7}$-$10^{-6}$~mbar. To speed up the
cleaning we warm the detector vessel from outside without exceeding
50~\textcelsius, mainly not to damage the photocathode of the PMT.
Moreover the argon filling procedure is
crucial: when a sufficient vacuum is achieved the detector is filled
with 99.9999~\% pure argon gas and the recirculation system is turned
on while cooling the detector. This procedure has two advantages: the
outgassed molecules are trapped into the getter cartridge during the
recirculation and the presence of argon gas inside the detector favors
uniform and fast cooling of all the the parts.
Drift electron lifetimes of some hundreds of microseconds have been
achieved corresponding to an oxygen equivalent impurity contamination
of ppb or less. 

\section{Operation in liquid argon and in double phase}
\label{sec:LEM-TPCOperation}
\begin{figure}[t]
\centering
\includegraphics[width=0.95\linewidth]{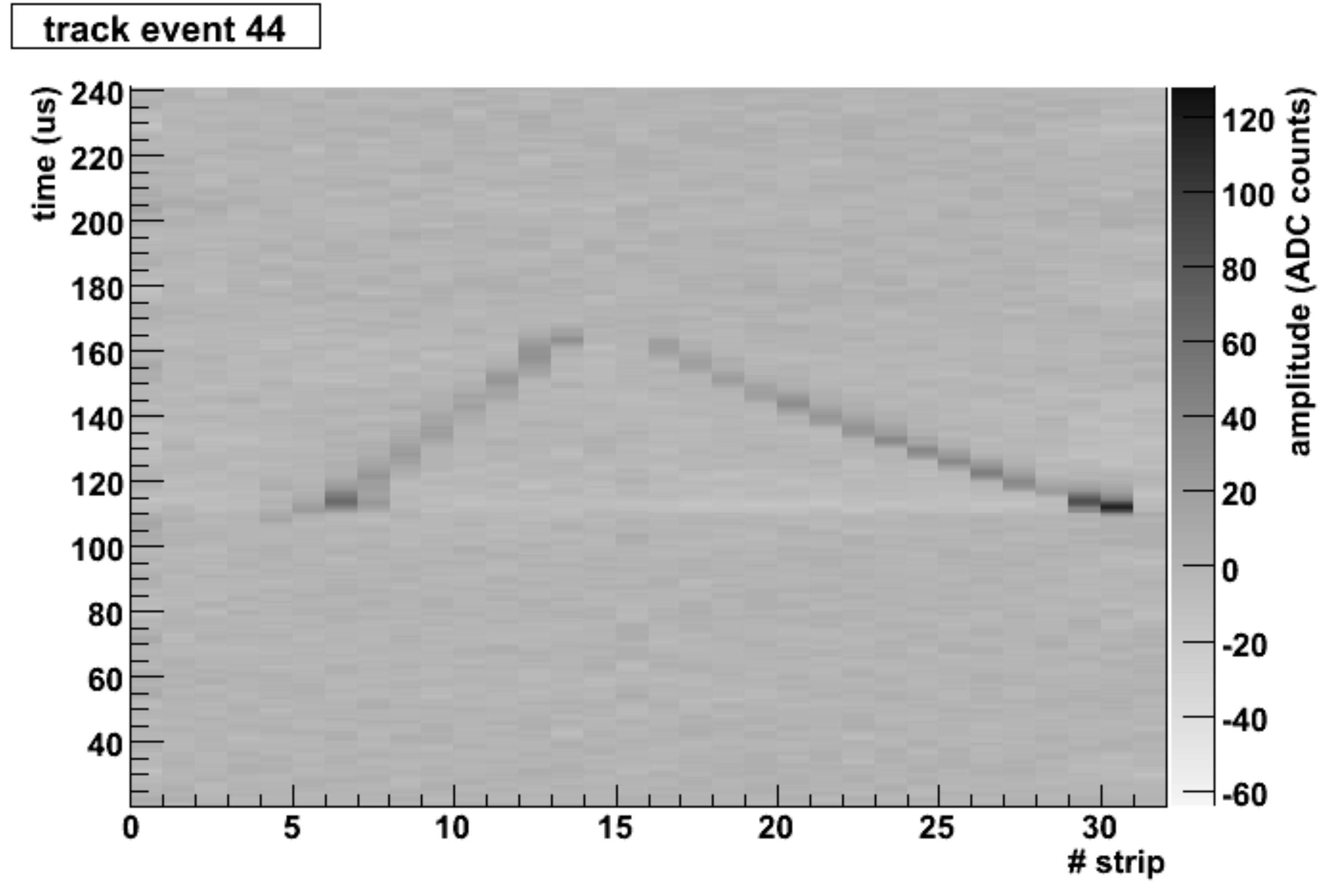}
\includegraphics[width=0.95\linewidth]{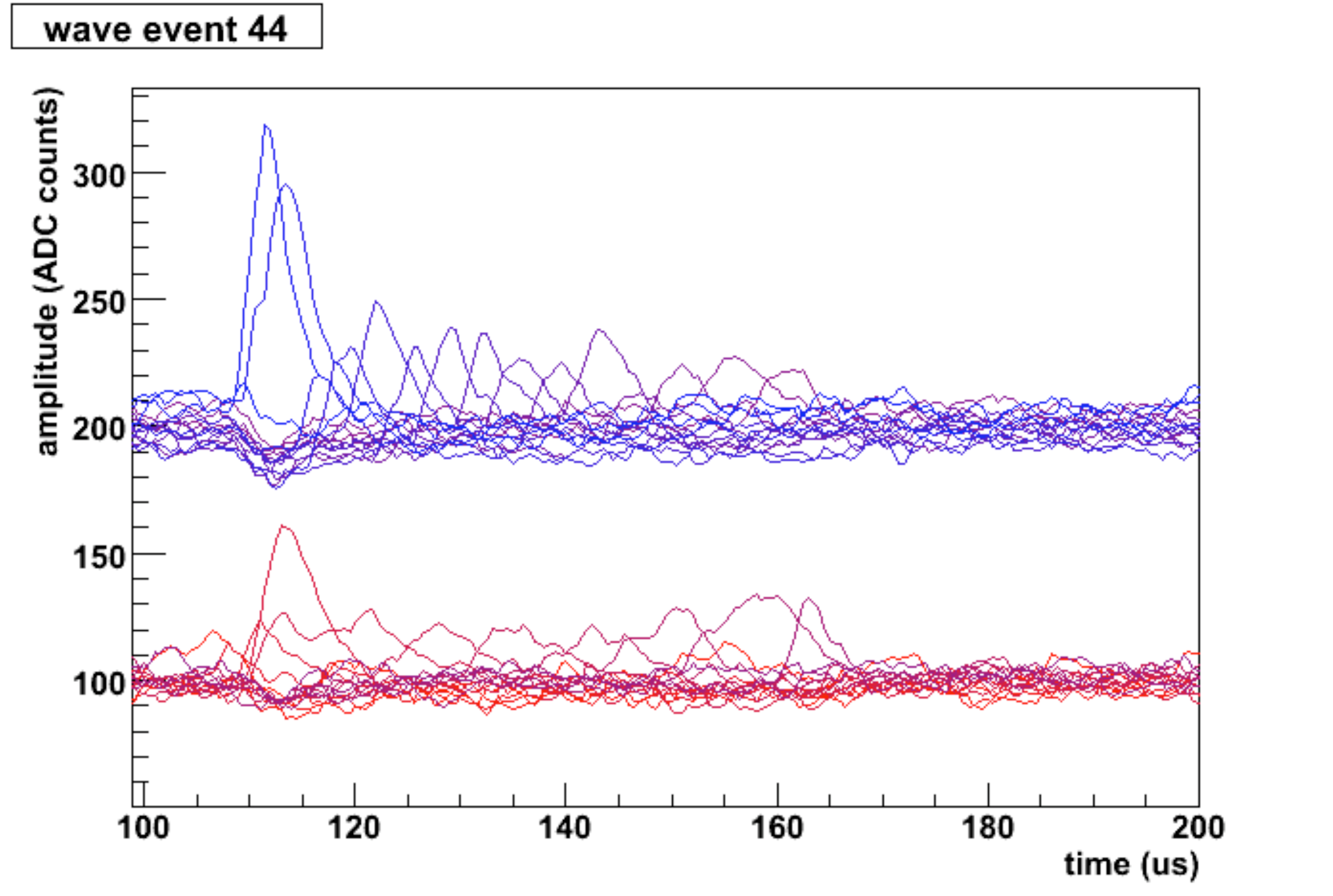}
\caption{Typical cosmic muon track with LEMs immersed in liquid
  (gain~1).\newline
  Top: the gray scale is proportional to the signal amplitude, the
  channel number (0-15 anode view and 16-31 second LEM top electrode
  view) is reported on the x-axis and the arrival time of the ionizing
  event on the y-axis.\newline
  Bottom: the 32 waveforms are plotted, the
  signals of the LEM plane on top and the ones of the anode below.}
\label{fig:liquidEvent}
\end{figure}

Picture~\ref{fig:liquidEvent} shows a cosmic muon track recorded
with the LEMs and the anode completely immersed in liquid: the
Townsend multiplication mechanism does not occur in this case given
the considered electric fields. 

A simple digital filter was applied in order to reduce a coherent
noise induced by the cathode power supply. The filter subtract the average
signal of all channels to each waveform. The signal to noise
ration after the noise reduction is about 80/5.

The aim of this measurement is to test the transparency of the
amplification system.

The electric field configuration is summarized in
table~\ref{tab:fieldConfig} together with the field configuration of
double phase operation.

\begin{table}
  \begin{tabular}{lrrr}
    \hline
    & & \multicolumn{2}{c}{$\Delta$V/x (kV/cm)} \\
    & distances (mm) & liquid & double phase \\
    \hline
    Anode-LEM$_2$ & 2 & 2.1 & 2.1 \\
    LEM$_2$ & 1.6 & 27 & 26 \\
    LEM$_2$-LEM$_1$ & 2 & 1.7 & 1.5 \\
    LEM$_1$ & 1.6 & 27 & 26 \\
    Drift & 95 & 0.8 & 0.7 \\
    \hline
  \end{tabular}
  \caption{Electric field configuration for liquid and double phase operation.}
  \label{tab:fieldConfig}
\end{table}

In double phase operation the liquid argon surface is positioned
between the two extraction grids in a 2.8~kV/cm electric field.
Electron amplification occurs in cold argon vapour ($\sim$1~bar and
87~K), about 3.4 times denser than at STP.

\begin{figure}[t]
\centering
\includegraphics[width=0.95\linewidth]{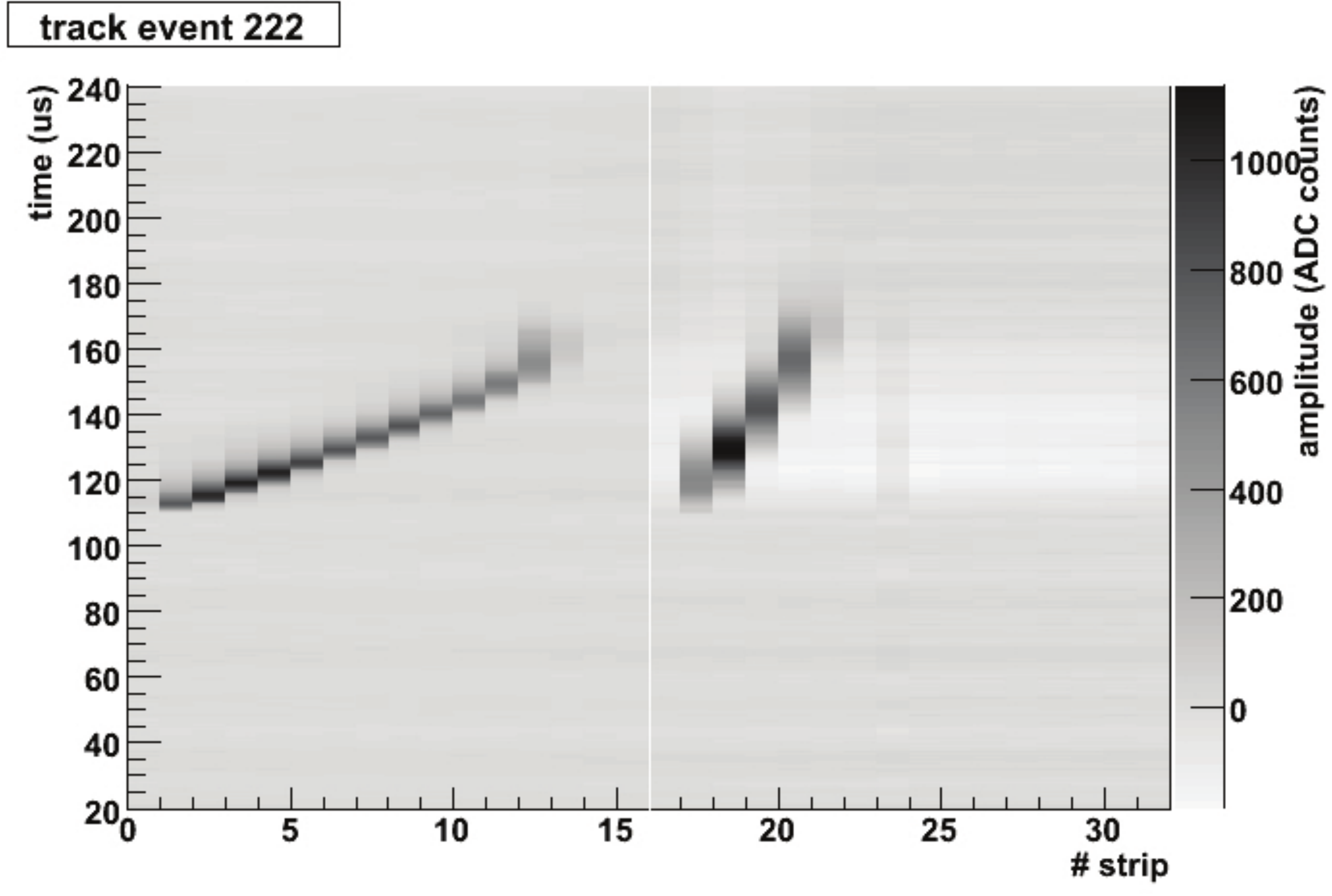}
\includegraphics[width=0.95\linewidth]{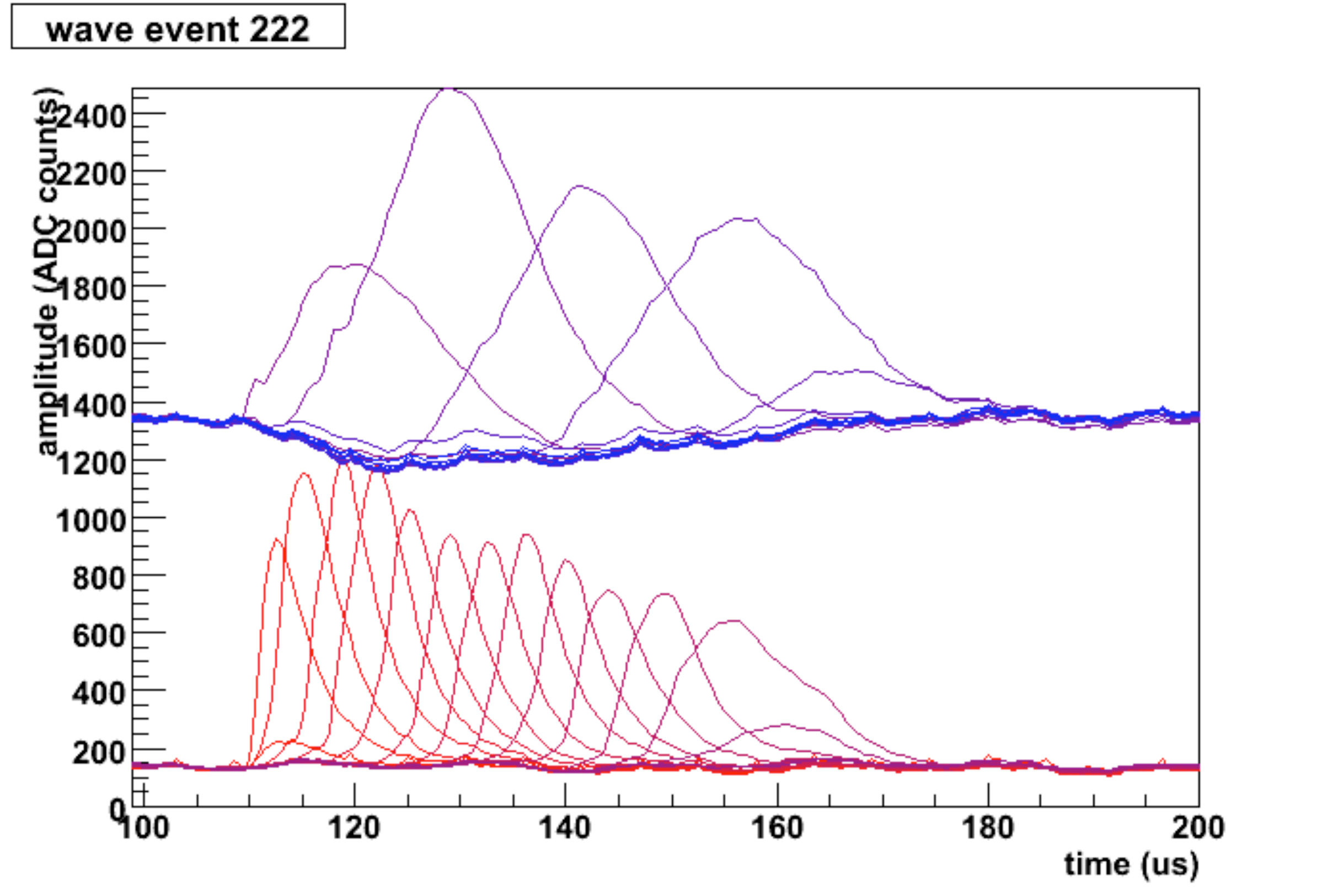}
\caption{Typical cosmic muon track in double phase operation.\newline
  Top: the gray scale is proportional to the signal amplitude, the
  channel number (0-15 anode view and 16-31 second LEM top electrode
  view) is reported on the x-axis and the arrival time of the ionizing
  event on the y-axis.\newline
  Bottom: the 32 waveforms are plotted, the
 signals of the LEM plane on top and the ones of the anode below.}
\label{fig:doublePhaseEvent}
\end{figure}

The figure~\ref{fig:doublePhaseEvent} shows a typical cosmic ray long
track. In this case no noise reduction was applied, because the
amplification improves considerably the signal to noise ratio
(S/N~=~800/10).
The base line distortion apparent on
the LEM electrode signals is not due to a failure or cross-talk
of the electronics. We interpret it as a capacitive pickup on
the upper LEM electrode of the physical signals induced on
the lower LEM face, which could be cured by connecting the
lower LEM face to a filter capacitor.

From the average energy loss of cosmic muons we evaluate the effective
gain of the LEM-TPC to be about 10: a minimum ionizing muon in liquid
argon at 0.5-1~kV/cm electric field releases about 10~fC/cm of
drifting electrons and on average the collected charge on the anode is
100~fC/cm.
It is evident the improvement of the signal to noise ratio between the
liquid operation and the double phase operation. In addition a
stable gain of~10 can compensate the degradation of the signal due
to impurities for very long drift paths. This fact makes the LEM-TPC
an interesting device for very long drift, cost-effective, liquid
argon experiments for neutrino physics and proton decay search.

\section{Conclusions}
We operated a novel kind of liquid argon TPC based on a
segmented LEM readout system, that allows charge amplification in
argon vapour.

We discussed in this paper the differences between the operation of the
LEM-TPC in single phase (LEMs and anode immersed in LAr), where no
gain is achievable, and the double phase operation (liquid-vapour),
where the gain was set to about 10.
From the cosmic muon tracks we evaluate a considerable improvement in
the signal to noise ratio.

Additional tests are needed to investigate the long term stability of the pure
argon double phase LEM-TPC and the maximum stable gain achievable.
A gain of 10 can compensate for the attenuation of the collected
charge for drifts of the order of 10~m.

For direct Dark Matter search higher gains ($\gtrapprox$~100) are
needed to reach an energy threshold down to tens of keV for nuclear
recoil detection.

\end{document}